\documentclass[sigconf, anonymous=false]{acmart}
\usepackage{figsize}
\usepackage{amsmath}
\usepackage{bbm}
\usepackage{todonotes}
\renewcommand\footnotetextcopyrightpermission[1]{}
\AtBeginDocument{%
  \providecommand\BibTeX{{%
    \normalfont B\kern-0.5em{\scshape i\kern-0.25em b}\kern-0.8em\TeX}}}

\newcommand{\algname}[1] {{\fontfamily{cmtt}\selectfont {#1}}}

\usepackage[absolute]{textpos}

\begin{document}

 \begin{textblock}{10}(3,0.4)
 \noindent\small  \begin{center}International Workshop on Industrial Recommendation Systems \\ in conjunction with ACM KDD 2020\end{center}
 \end{textblock}
\title{Multi-sided Exposure Bias in Recommendation}

\author{Himan Abdollahpouri}
\email{himan.abdollahpouri@colorado.edu}
\affiliation{%
  \institution{University of Colorado Boulder}
  \city{Boulder}
  \country{USA}
}
\author{Masoud Mansoury}
\authornote{This author also has affiliation in School of Computing, DePaul University, Chicago, USA, mmansou4@depaul.edu.}
\email{m.mansoury@tue.nl}
\affiliation{%
  \institution{Eindhoven University of Technology}
  \city{Eindhoven}
  \country{The Netherlands}
}

\renewcommand{\shortauthors}{Abdollahpouri and Mansoury}

\begin{abstract}
  Academic research in recommender systems has been greatly focusing on the accuracy-related measures of recommendations. Even when non-accuracy measures such as popularity bias, diversity, and novelty are studied, it is often solely from the users' perspective. However, many real-world recommenders are often multi-stakeholder environments in which the needs and interests of several stakeholders should be addressed in the recommendation process. In this paper, we focus on the popularity bias problem which is a well-known property of many recommendation algorithms where few popular items are over-recommended while the majority of other items do not get proportional attention and address its impact on different stakeholders. Using several recommendation algorithms and two publicly available datasets in music and movie domains, we empirically show the inherent popularity bias of the algorithms and how this bias impacts different stakeholders such as users and suppliers of the items. We also propose metrics to measure the exposure bias of recommendation algorithms from the perspective of different stakeholders.

\end{abstract}




\maketitle

\section{Introduction}
Popularity bias is a well-known phenomenon in recommender systems: popular items are recommended even more frequently than their popularity would warrant, amplifying the long-tail effect already present in many recommendation domains. Prior research has examined the impact of this bias on some properties of the recommenders such as aggregate diversity (aka catalog coverage) \cite{Vargas:2011:RRN:2043932.2043955,adomavicius2011improving}. One of the consequences of the popularity bias is disfavoring less popular items where the recommendations are not fair in terms of the amount of exposure they give to different items with varying degree of popularity: an exposure bias. However, as we discuss in \cite{abdollahpourimultistakeholder2020}, many recommender systems are multi-stakeholder environments in which the needs and interests of multiple stakeholders should be taken into account in the implementation and evaluation of such systems. 

In many multi-stakeholer recommenders as described in \cite{abdollahpourimultistakeholder2020} two main stakeholders (or what often is being referred to as \textit{sides} in multi-sided platforms \cite{hagiu2015multi} ) can be identified: consumers (aka users) and suppliers. For instance, in a music platform such as Spotify, on one side there are users who get recommendations for songs in which they are interested and, on the other side, there are artists whose songs are being recommended to different users. The popularity bias can be investigated from both sides' perspective.

Regarding the users, not everyone has the same level of interest in popular items. In the music domain as an example, some users might be interested in internationally popular artists such as Drake, Beyonc\'{e}, or Ed Sheeran and some might be more interested in artists from their own culture that might not necessarily have the same popularity as the aforementioned artists (such as the Iranian musician Kayhan Kalhor) or generally they prefer certain type of music that might not be popular among the majority of other users (such as country music). With that being said, we expect the personalization to handle this difference in taste but as we will see in section ~\ref{user_impact} that is certainly not the case. 

The suppliers also do not have the same level of popularity. In many recommendation domains including movies, music, or even house sharing, few suppliers have a large audience while the majority of others may not be as popular though they still might have their fair share of audience. Now the question is, do recommender systems let different suppliers with varying degree of popularity reach their desired audience? Again, the short answer is No as we will see more details in section ~\ref{supplier_impact}.

Investigating the impact of recommendation algorithms on the exposure bias on both users and suppliers is the focus of this paper. We study several recommendation models in terms of their inherent popularity bias and propose metrics that can measure such impact.

\section{Experimental Setting}
\subsection{Data}
 We have used two publicly available datasets for our experiments. We needed datasets that either had information about the supplier of the items or we could extract them. We found two: the first one is a sample of the Last.fm (LFM-1b) dataset \cite{schedl2016lfm} used in \cite{dominik2019unfairness}. The dataset contains user interactions with songs (and the corresponding albums). We used the same methodology in \cite{dominik2019unfairness} to turn the interaction data into rating data using the frequency of the interactions with each item (more interactions with an items will result in higher ratings). In addition, we used albums as the items to reduce the size and sparsity of the item dimension, therefore the recommendation task is to recommend albums to users. We considered the artists of each album as the supplier. Each album is associated with an artist. Artists could have multiple albums. We removed users with less than 20 ratings so only consider users for which we have enough data. The resulting dataset contains 274,707 ratings by 2,697 users to 6,006 albums from 1,998 artists.
 
 The second dataset is the MovieLens 1M dataset\footnote{Our experiments showed similar results on MovieLens 20M, and so we continue to use ML1M for efficiency reasons.}. This dataset does not have the information about the suppliers. We considered the director of each movie as the supplier of that movie and we extracted that information from the IMDB API. Total number of ratings in the MovieLens 1M data is 1,000,209 given by 6,040 users to 3,706 items. Overall, we were able to extract the director information for 3,043 movies reducing the ratings to 995,487. The total number of directors is 831. 
 
 We used 80\% of each dataset as our training set and the other 20\% for the test.
 
 \subsection{Algorithms}
 For our analysis, we have used three personalized recommendation algorithms: \textit{Biased Matrix Factorization} (\algname{Biased-MF}) \cite{koren2009matrix}, \textit{User-based Collaborative Filtering} (\algname{User-CF}) \cite{aggarwal2016neighborhood}, and \textit{Item-based Collaborative Filtering} (\algname{Item-CF}) \cite{deshpande2004item}. All algorithms are tuned to achieve their best performance in terms of precision. The size of the recommendation lists for each user is set to 10. We also include a non-personalized \algname{Most-popular} algorithm which only recommends the 10 most popular items to every user given they have not rated the items before. We used LibRec \cite{Guo2015} and \textit{librec-auto} \cite{mansoury2018automating} for running the algorithms.

\section{Popularity Bias}
Skew in wealth distribution is well-known: The richest 10\% of adults in the world own 85\% of global household wealth while the bottom half collectively owns barely 1\% \footnote{https://www.wider.unu.edu/publication/global-distribution-household-wealth}; in recommender systems a similar problem exists: a small number of popular items appear frequently in user profiles and a much larger number of less popular items appear rarely. This bias can originate from two different sources: data and algorithms. 

\subsection{Bias in Data}

\begin{figure}
\centering
\SetFigLayout{3}{2}
 \subfigure[MovieLens]{\includegraphics[width=1.6in]{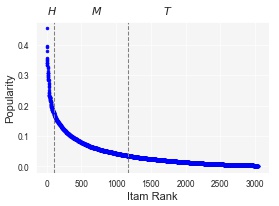}}
\subfigure[Last.fm]{\includegraphics[width=1.6in]{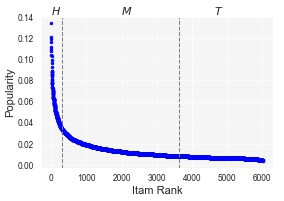}}
  \hfill
\caption{The Long-tail Distribution of Rating data. $H$ items are a small number of very popular items that take up around 20\% of the entire ratings. $T$ is the larger number of less popular items which collectively take up roughly 20\% of the ratings, and $M$ includes those items in between that receive around 60\% of the ratings, collectively.} \label{longtail}
\end{figure}

Rating data is generally skewed towards more popular items. Figure ~\ref{longtail} shows the percentage of users who rated different items in MovieLens and Last.fm datasets: the popularity of each item. Items are ranked from the most popular to the least with the most popular item being on the far left on the x-axis. Three different groups of items can be seen in these plots: $H$ which represents few items that are very popular and take up around 20\% of the entire ratings according to the Pareto Principle \cite{sanders1987pareto}. $T$ is the larger number of less popular items which collectively take up roughly 20\% of the ratings, and $M$ includes those items in between that receive around 60\% of the ratings, collectively. The curve has a long-tail shape \cite{anderson2006long,celma2008hits} indicating few popular items are taking up the majority of the ratings while many other items on the far right of the curve have not received much attention. This type of distribution can be found in many other domains such as e-commerce where few products are best-sellers, online dating where few profiles attract the majority of the attention, social networking platforms where few users have millions of followers, to name a few. 

The bias in rating data could be due to two different reasons:
\begin{itemize}
    \item \textbf{External Bias:} Some items and products are inherently more popular than others even outside of the recommender systems and in the real world. For instance, even before the music streaming services emerge, there were always few artists that were nationally or internationally popular such as \textit{Shakira}, \textit{Jennifer Lopez}, or \textit{Enrique Iglesias}. As a result of this external bias (or tendency) towards popular artists, users also often listen to those artists more on streaming services and hence they get more user interactions. 
    \item \textbf{Feedback Loop}: Since the recommendation algorithms have a higher tendency towards recommending popular items, these items have a higher chance to be recommended to the users and hence garnering a larger number of interactions from the users. When these interactions are logged and stored, the popularity of those items in the rating data increases since they get more and more interactions over time \cite{jiang2019,sinha2016}.  
\end{itemize}

\subsection{Bias in Algorithm}\label{bias_in_alg}
Due to this imbalance property of the rating data, often algorithms inherit this bias and, in many cases, intensify it by over-recommending the popular items and, therefore, giving them a higher opportunity of being rated by more users: \textit{the rich get richer and the poor get poorer} \cite{abdollahpouri2017controlling}.

\begin{figure}
\centering
\SetFigLayout{2}{1}
 \subfigure[MovieLens]{\includegraphics[width=3.3in]{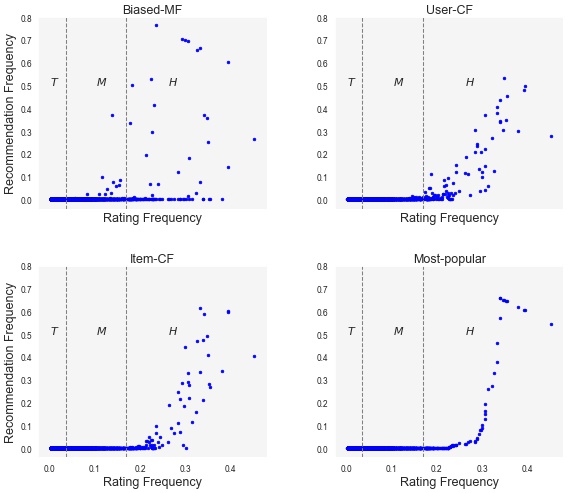}\label{corr_scatter_movielens}}
  \hfill
 \subfigure[Last.fm]{\includegraphics[width=3.3in]{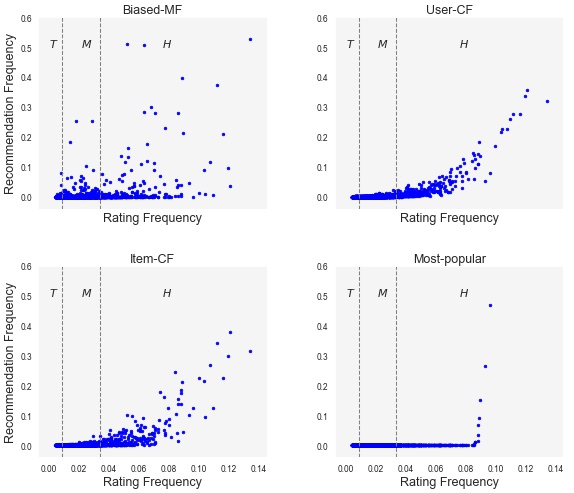}\label{corr_scatter_lastfm}}
  \hfill
\caption{Item popularity versus recommendation popularity} \label{corr_scatter}
\end{figure}

Figure ~\ref{corr_scatter} shows the percentage of users who have rated an item on the x-axis (the popularity of an item in the data) and the percentage of users who received that item in their recommendations using four different recommendation algorithms \algname{Biased-MF}, \algname{User-CF}, \algname{Item-CF}, and \algname{Most-popular} in both datasets. The plots aim at showing the correlation between the popularity of an item in the rating data versus how often it is recommended to different users. It is clear that in all four algorithms, many items are either never recommended or just rarely recommended. Among the three personalized algorithms, \algname{Item-CF} and \algname{User-CF} show the strongest evidence that popular items are recommended much more frequently than the others. In fact, they are recommended to a much greater degree than even what their initial popularity warrants. For instance, the popularity of some items have been amplified from roughly 0.4 to 0.7 indicating a 75\% increase. Both \algname{Item-CF} and \algname{User-CF} are over-promoting popular items (items on the right side of the x-axis) while significantly hurting other items by not recommending them proportionate to what their popularity in data warrants. In fact, the vast majority of the items on the left are never recommended indicating an extreme bias of the algorithms towards popular items and against less popular ones. \algname{Biased-MF} does not show a positive correlation between popularity in data and in recommendations although some items are still over-recommended (have much higher popularity in recommendations versus what they had in rating data). However, this over-recommendation is not concentrated on only popular items and some items from lower popularity values are also over-recommended. \algname{Most-popular} obviously shows the strongest bias but, unlike the other three, it is not a personalized method\footnote{If an item in the top 10 is already rated by a user it will be replaced by another popular item. That is why there is an inflection point in the scatter plot for this algorithm on MovieLens dataset.}.

\section{Multi-sided Exposure Bias}
We measure the impact of popularity bias on different stakeholders in terms of exposure: how the bias in algorithms prevents users to be exposed to the appropriate range of items and also how it stops items from different suppliers to be exposed to their desired audience. 

\begin{figure}
\centering
\SetFigLayout{2}{1}
 \subfigure[MovieLens]{\includegraphics[width=2.5in]{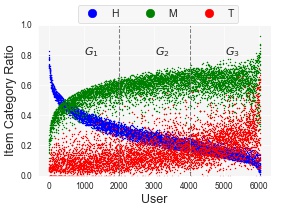}}
  \hfill
 \subfigure[Last.fm]{\includegraphics[width=2.5in]{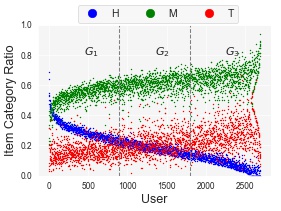}}
  \hfill
\caption{Users' Propensity towards item popularity} \label{user_propensity}
\end{figure} 

\subsection{Exposure Bias From the Users' Perspective}\label{user_impact}

\begin{figure}
\centering
\SetFigLayout{2}{1}
 \subfigure[MovieLens]{\includegraphics[width=3.5in]{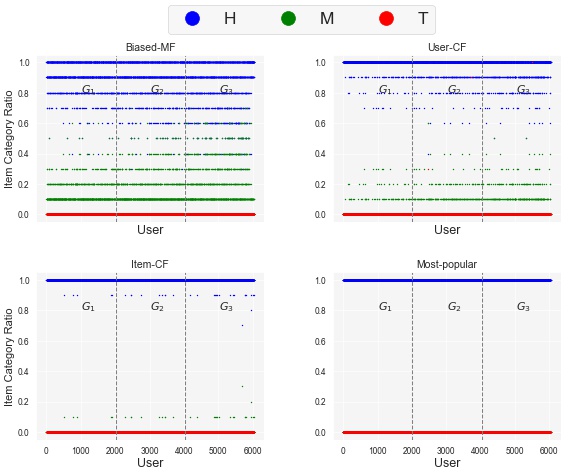}}
  \hfill
 \subfigure[Last.fm]{\includegraphics[width=3.5in]{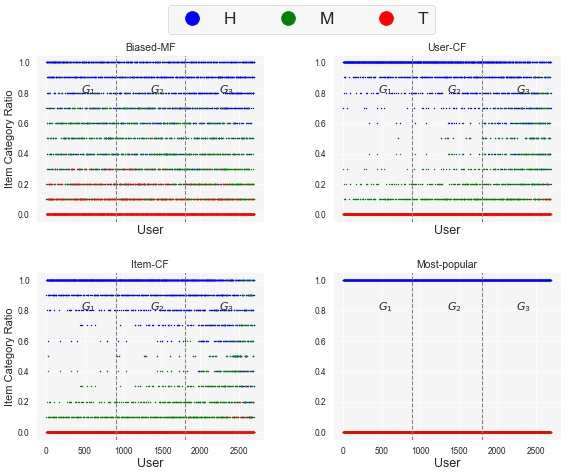}}
  \hfill
\caption{Users' centric view of popularity bias} \label{ms_impact-users}
\end{figure} 

 Not every user is equally interested in popular items. In cinema, for instance, some might be interested in movies from Yasujiro Ozu, Abbas Kiarostami, or John Cassavetes, and others may enjoy more mainstream directors such as James Cameron or Steven Spielberg. Figure ~\ref{user_propensity} shows the ratio of rated items for three item categories $H$, $M$, and $T$ in the profiles of different users in the MovieLens 1M and Last.fm datasets. Users are sorted from the highest interest towards popular items to the least and divided into three equal-sized bins $G=\{G_1$,$G_2$,$G_3\}$ from most popularity-focused to least. The y-axis shows the proportion of each user's profile devoted to different popularity categories. The narrow blue band shows the proportion of each users profile that consists of popular items (i.e. the $H$ category), and its monotone decrease reflects the way the users are ranked. Note, however, that all groups have rated many items from the middle (green) and tail (red) parts of the distribution.

The plots in Figure ~\ref{ms_impact-users} are parallel to Figure ~\ref{user_propensity}, with the users ordered by their popularity interest, but now the y-axis shows the proportion of recommended items using different algorithms from different item popularity categories. The difference with the original user profiles in rating data especially in the case of \algname{Most-popular}, \algname{Item-CF}, and \algname{User-CF} is stark where the users' profiles are rich in diverse popularity categories, the generated recommendations are nowhere close to what the user has shown interest at. In fact, in \algname{Item-CF} almost 100\% of the recommendations are from the head category, even for the users with the most niche-oriented profiles. Tail items do not appear at all. We demonstrated here that popularity bias in the algorithm is not just a problem from a global, system, perspective. It is also a problem from the user perspective \cite{himan2019b}: users are not getting recommendations that reflect the diversity of their profiles and the users with the most niche tastes ($G_3$) are the most poorly served. 

To measure the impact of popularity bias on users, we need to compare two lists together: the list of the items in a users' profile and the list of the items recommended to the user. To do this comparison, we need to compute a discrete probability distribution $P$ for each user $u$, reflecting the popularity of the items found in their profile $\rho_u$ over each item category $c \in C$ 
(in this paper, $C=\{H,M,T\}$). We also need a corresponding distribution $Q$ over the recommendation list given to user $u$, $\ell_u$, indicating what popularity categories are found among the list of recommended items. In Steck \cite{steck2018calibrated} and Kaya et al. \cite{kaya2019comparison} where, unlike this paper, the calibration is done according to genres, the rating data is binarized to reflect just ``liked'' items. We are retaining the original ratings when computing over user profiles and, instead, using Vargas et al.'s \cite{vargas2013exploiting} measure of category propensity which has the users' rating component in it. Note that unlike the genre labels in \cite{steck2018calibrated} where it is possible for a movie to have multiple genres, each item only has one popularity category. 

\begin{equation}
p(c|u)=\frac{\sum_{i \in \rho_u}r(u,i)\mathbbm{1}(i \in c)}{\sum_{c_j \in C}\sum_{i \in \rho_u }r(u,i)\mathbbm{1}(i \in c_j)}
\end{equation}  

\begin{equation}
q(c|u)=\frac{\sum_{i \in \ell_u}\mathbbm{1}(i \in c)}{\sum_{c_j \in C}\sum_{i \in \ell_u }\mathbbm{1}(i \in c_j)}
\end{equation} 

$\mathbbm{1}(.)$ is the indicator function returning zero when its argument is False and 1 otherwise. Once we have $P$ and $Q$, we can measure the distance using Jensen–Shannon divergence, which is a modification of KL-Divergence that has two useful properties which KL-divergence lacks: 1) it is symmetric: $\mathfrak{J}(P,Q)=\mathfrak{J}(Q,P)$ and 2) it has always a finite value even when there is a zero in $Q$. For our application, it is particularly important that the function be well-behaved at the zero point since it is possible for certain item categories to have zero items in them in the recommendation list. 

\begin{equation}
    \mathfrak{J}(P,Q)=\frac{1}{2}KL(P,M)+\frac{1}{2}KL(Q,M), \; \; M=\frac{1}{2}(P+Q)
\end{equation}
\noindent where $KL$ is the KL divergence.

We introduce the following metric to quantify the exposure bias from the users' perspective. 

\textbf{Users' Popularity propensity Deviation (UPD):}
 
 Having different groups of users with varying degree of interest towards popular items, we measure the average deviation of the recommendations given to the users in each group in terms of item popularity. More formally,

\begin{equation}
    UPD=\frac{\sum_{\textsl{g} \in G}\frac{\sum_{u \in \textsl{g}}\mathfrak{J}(P(\rho_u),Q(\ell_u))}{|\textsl{g}|}}{|G|}
    \end{equation}
\noindent  where $|\textsl{g}|$ is the number of users in group $\textsl{g}$ and $|G|$ is the number of user groups. $UPD$ can be also seen as the average miscalibration of the recommendations from the perspective of users in different groups.
\subsection{Exposure Bias From the Suppliers' Perspective}\label{supplier_impact}
As noted above, multi-stakeholder analysis in recommendation also includes providers or as termed here \textit{suppliers}, ``those entities that supply or otherwise stand behind the recommended items''~\cite{abdollahpourimultistakeholder2020}. We can think of many different kinds of contributors standing behind a particular movie: for the purposes of this paper, we will focus on movie directors. In the music domain, often the artists are considered as the suppliers of the songs \cite{mehrotra2018towards}. In this paper, we also make the same assumption on Last.fm dataset. 

\begin{figure}
\centering
\SetFigLayout{2}{1}
 \subfigure[MovieLens]{\includegraphics[width=3.5in]{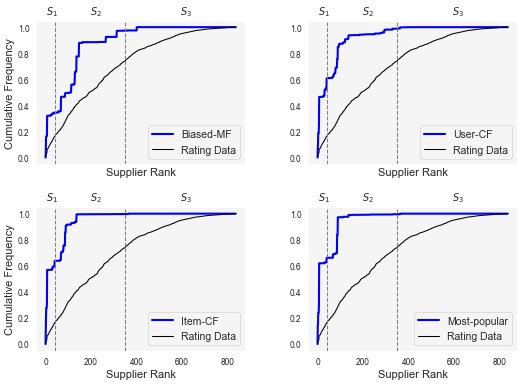}\label{fig:supplier_centric_movie}}
  \hfill
 \subfigure[Last.fm]{\includegraphics[width=3.5in]{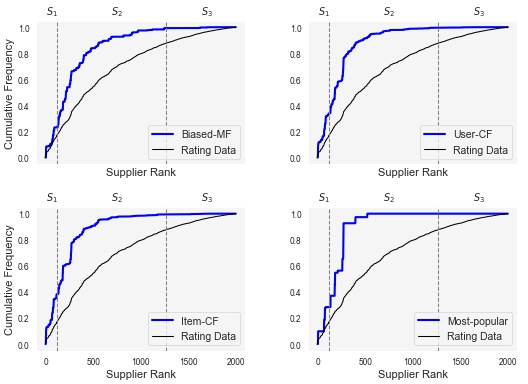}\label{fig:supplier_centric_lastfm}}
  \hfill
\caption{Suppliers' centric view of popularity bias} \label{fig:supplier_centric}
\end{figure}

 We create three supplier groups $S=\{S_1,S_2,S_3\}$: $S_1$ represents few popular suppliers whose items take up 20\% of the ratings, $S_2$ are larger number of suppliers with medium popularity whose items take up around 60\% of the ratings, and $S_3$ are the less popular suppliers whose items get 20\% of the ratings.
Figure ~\ref{fig:supplier_centric} shows the rank of different directors in MovieLens and artists in Lats.fm datasets by popularity and the corresponding recommendation results from different algorithms. The recommendations have amplified the popularity of the popular suppliers (the ones on the extreme left) while suppressing the less popular ones dramatically. Strikingly, using \algname{Item-CF}, movies from just 3 directors in $S_1$ (less than 0.4\% of the suppliers here) take up 50\% of the recommendations produced, while items items from the $S_3$ are seeing essentially zero recommendations. 

To quantify the impact of popularity bias on the supplier exposure, we measure the amount of deviation different groups of suppliers experience in terms of exposure. In other words, the degree of over-recommendation or under-recommendation of suppliers from different groups. 

\textbf{Supplier Popularity Deviation (SPD):}
\begin{equation}
    SPD=\frac{\sum_{s \in S}|q(s)-p(s)|}{|S|}
\end{equation}

\begin{equation*}
   q(s)=\frac{\sum_{u \in U}\sum_{i \in \ell_u}\mathbbm{1}(A(i) \in s)}{n\times|U|}, \; p(s)=\frac{\sum_{u \in U}\sum_{i \in \rho_u}\mathbbm{1}(A(i) \in s)}{\sum_{u \in U} |\rho_u|}
\end{equation*}

\noindent where $q(s)$ is the ratio of recommendations that come from items of supplier group $s$. $p(s)$ is the ratio of ratings that come from items of supplier group $s$. $U$ is the set of users and $A(.)$ is a mapping function that returns the supplier for each item. 

In fact, $1-SPD$ can be considered as \textit{Proportional Supplier Fairness} \cite{kusner2017counterfactual} since it measures how the items from different supplier groups are exposed to different users \textit{proportional} to their popularity in rating data.

\begin{figure*}
\centering
\SetFigLayout{2}{1}
 \subfigure[UPD vs SPD]{\includegraphics[width=2in]{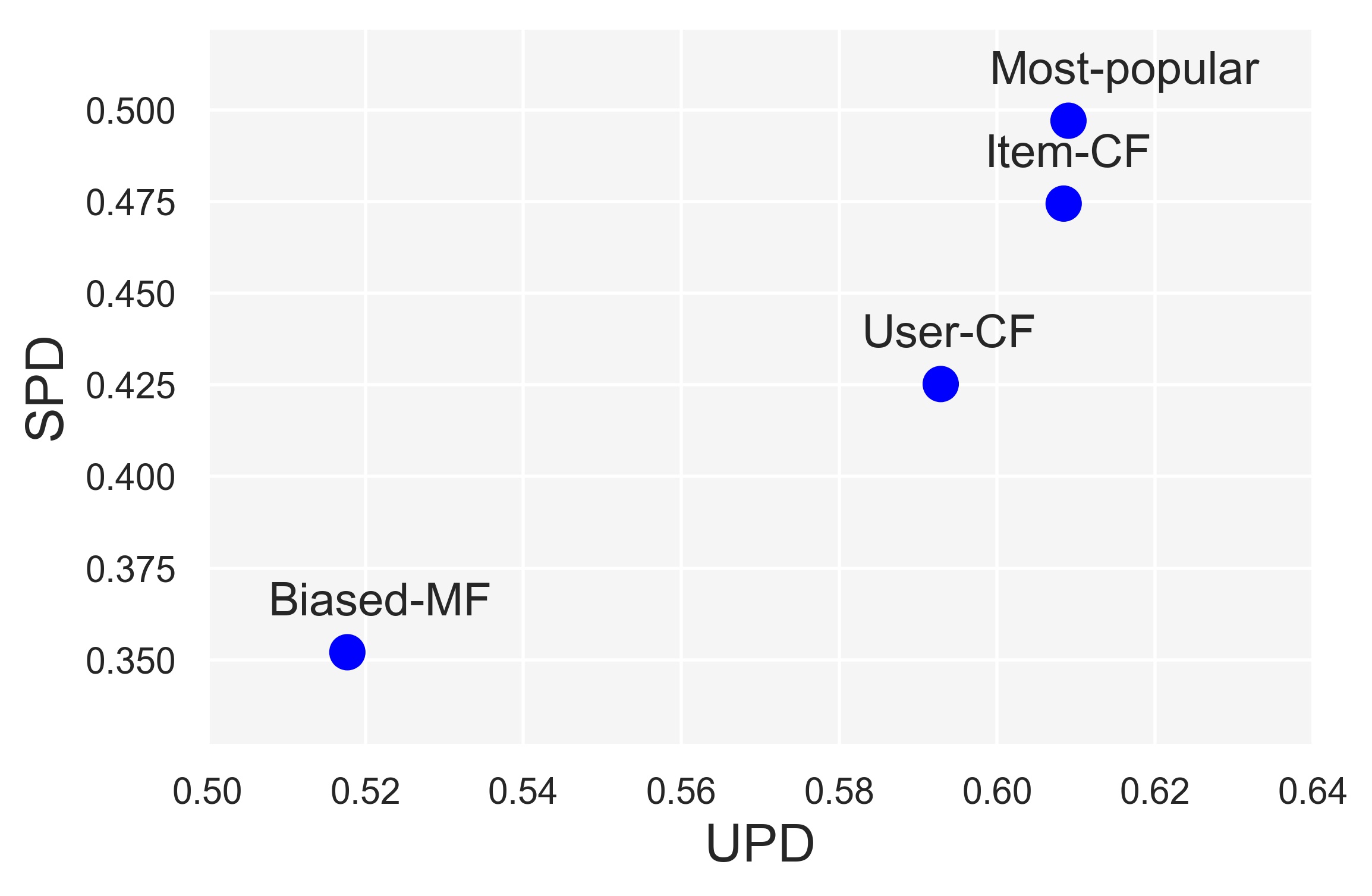}\label{fig:upd_spd_movie}}
 \subfigure[UPD vs Precision]{\includegraphics[width=2in]{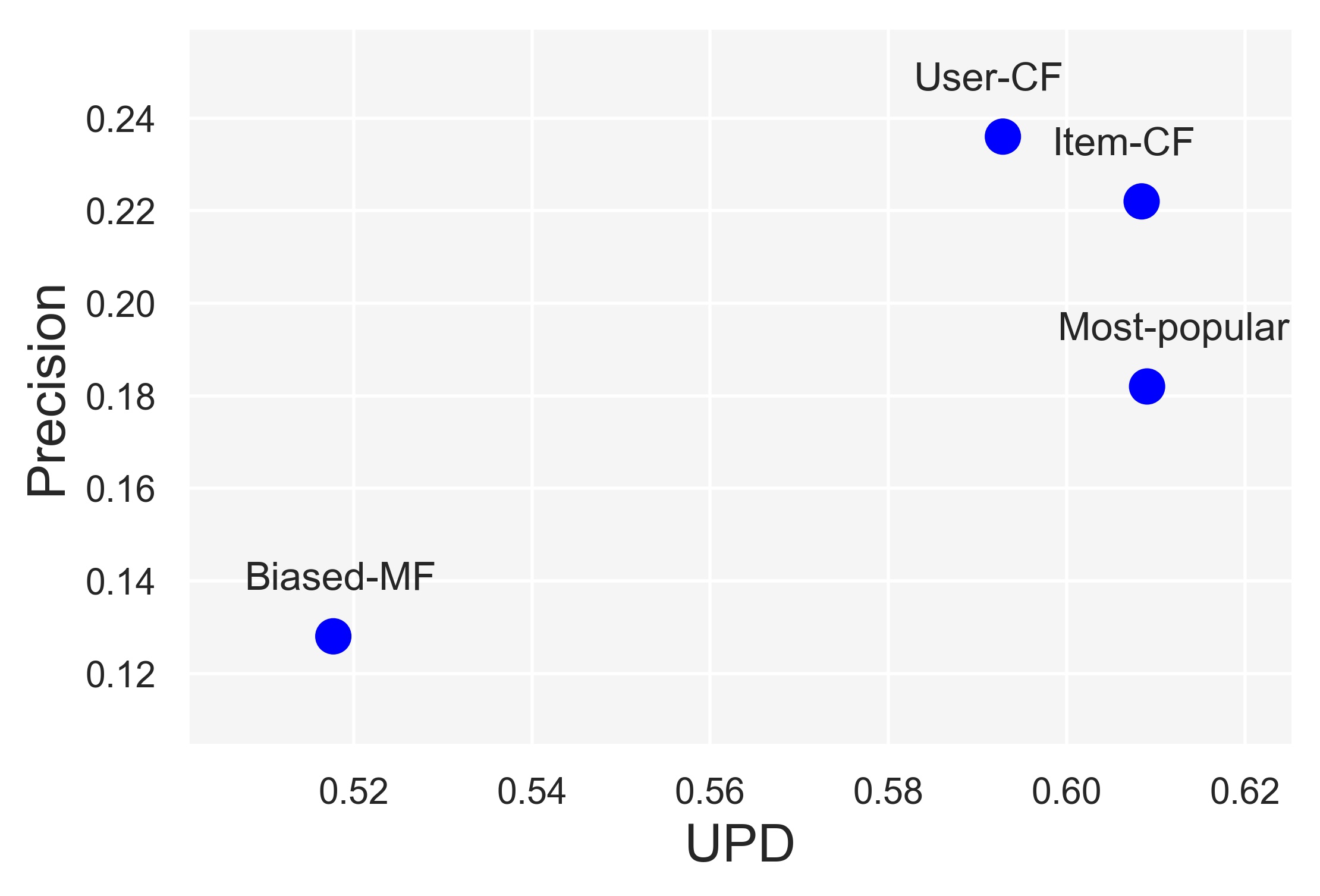}\label{fig:upd_prec_movie}}
      \subfigure[SPD vs Coverage]{\includegraphics[width=2in]{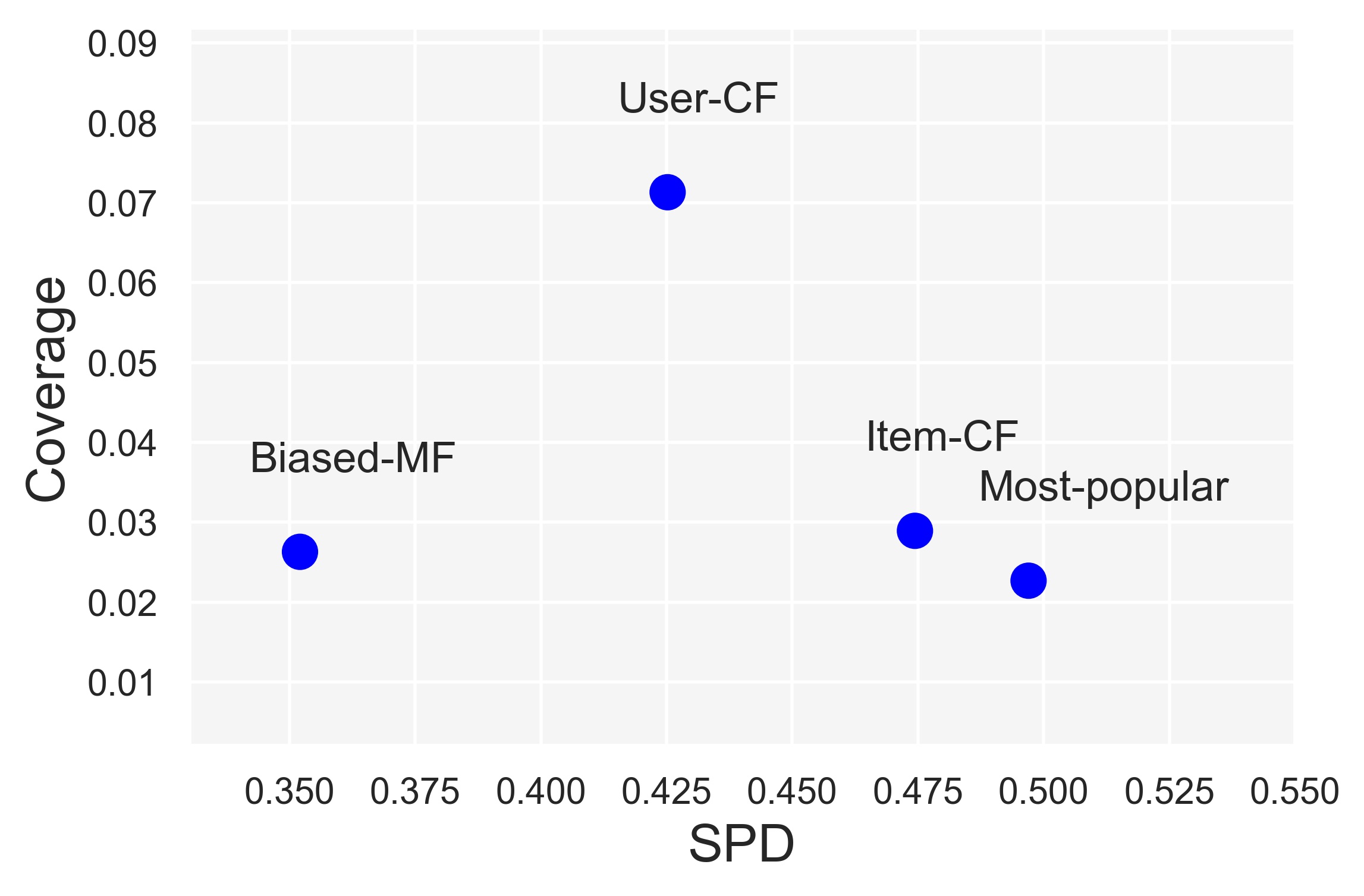}\label{fig:spd_cov_movie}}
\caption{The relationship between different metrics (MovieLens)} \label{fig:corss_movie}
\end{figure*} 

\begin{figure*}
\centering
\SetFigLayout{2}{1}
 \subfigure[UPD vs SPD]{\includegraphics[width=2in]{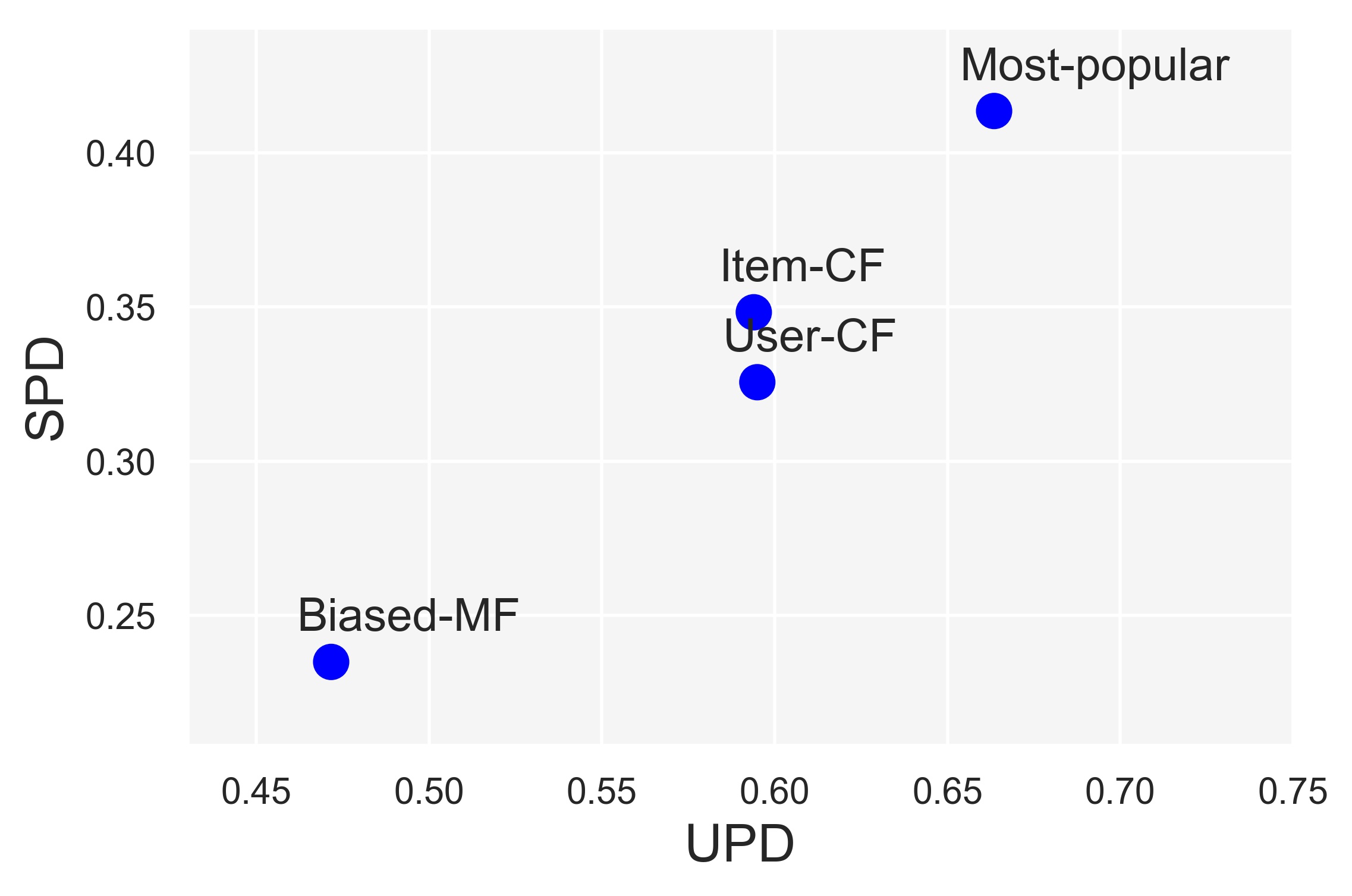}\label{fig:upd_spd_lastfm}}
 \subfigure[UPD vs Precision]{\includegraphics[width=2in]{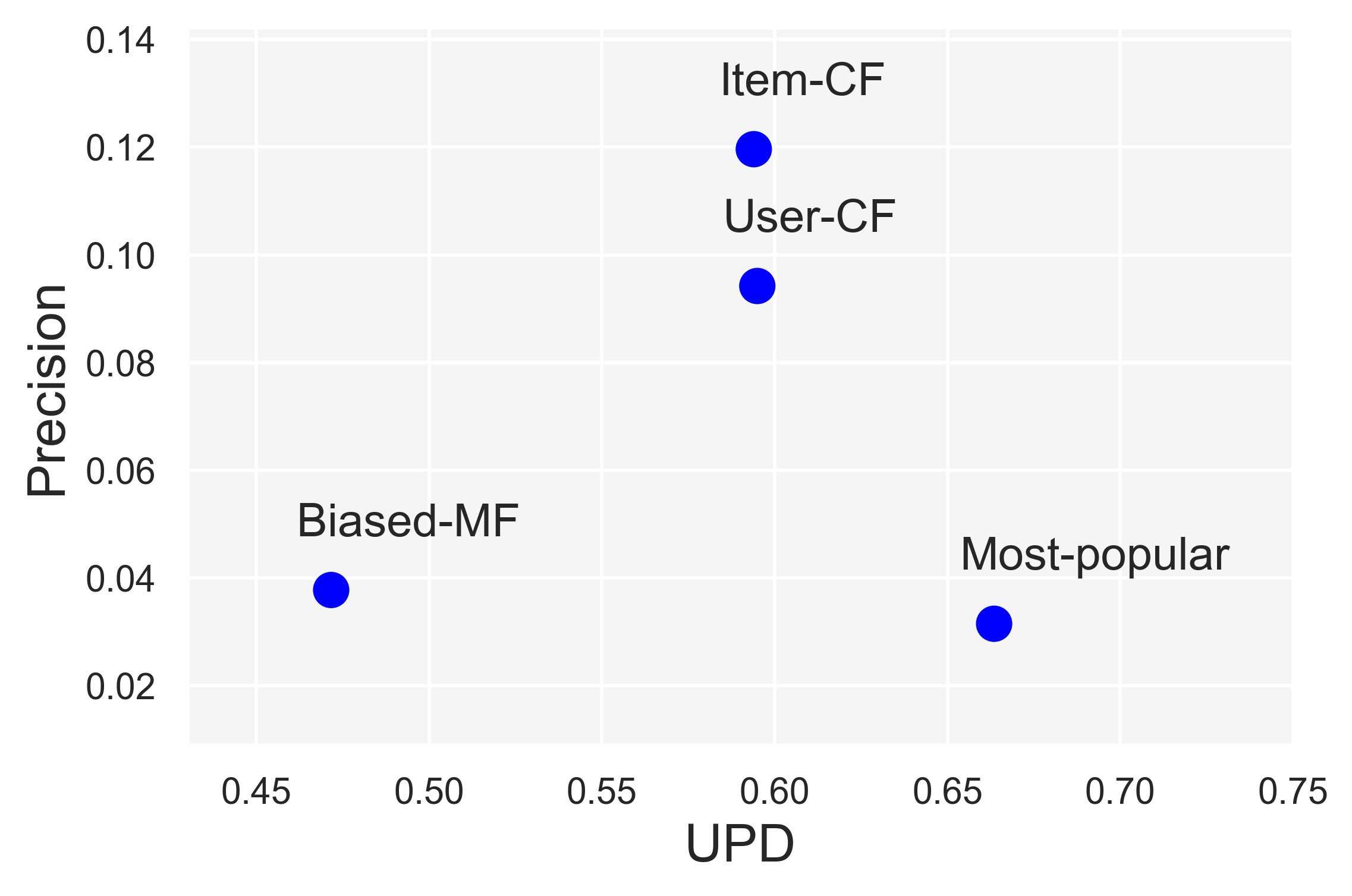}\label{fig:upd_prec_lastfm}}  
\subfigure[SPD vs Coverage]{\includegraphics[width=2in]{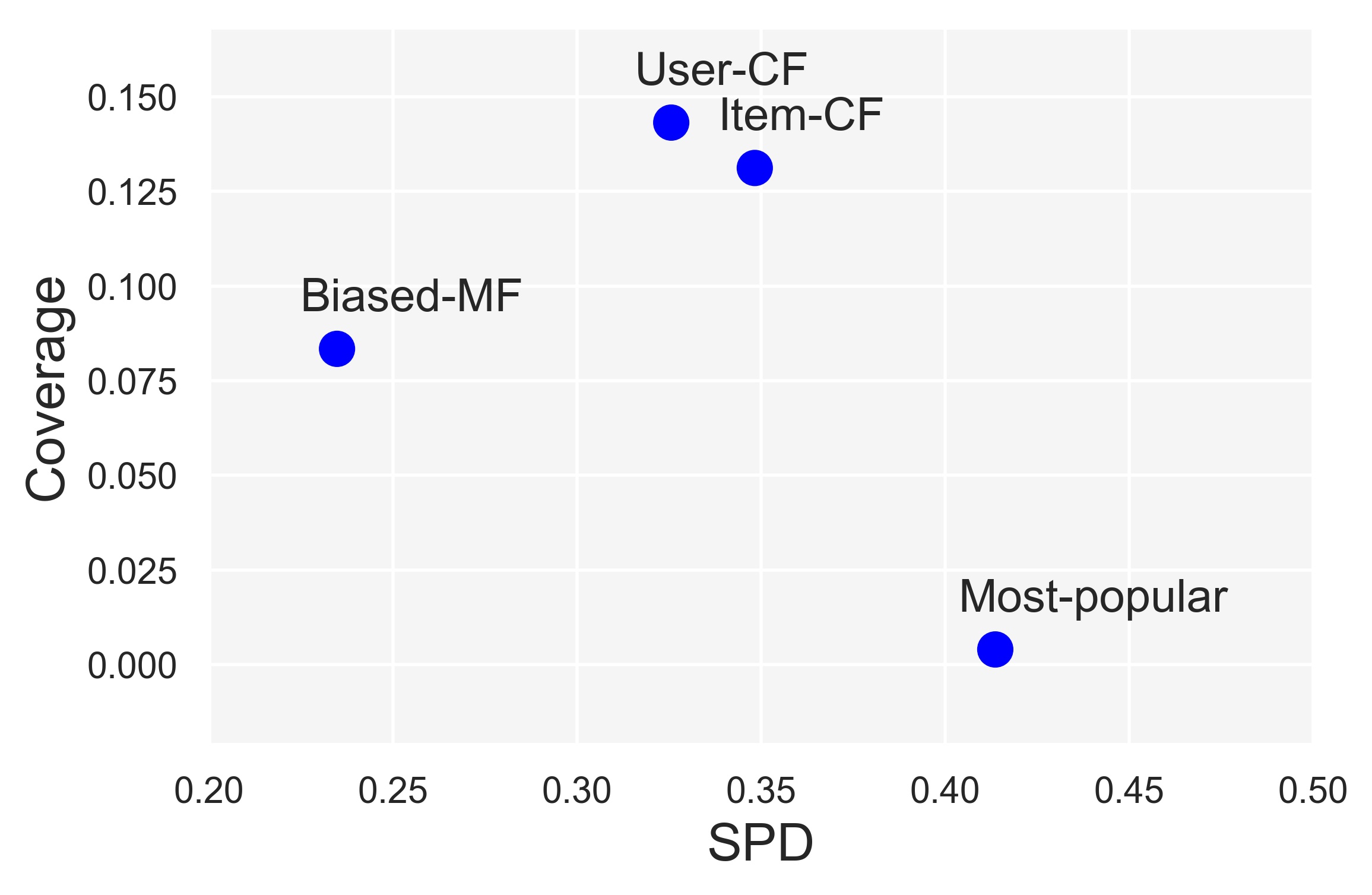}\label{fig:spd_cov_lastfm}}
\caption{The relationship between different metrics (Last.fm)} \label{fig:corss_lastfm}
\end{figure*}

\section{Discussion and Future Work} 
One important consideration regarding the deviation of popularity from the users' perspective ($UPD$) and suppliers' perspective ($SPD$) is how these two metrics behave with respect to one another. In other words, whether calibrating the recommendations for the users in terms of popularity would make the experience for the suppliers also better. Figures ~\ref{fig:upd_spd_movie} and ~\ref{fig:upd_spd_lastfm} show the connection between these two in the MovieLens and Last.fm datasets, respectively. We can see that, generally, the lower the $UPD$ (better calibration from the users' perspective) the lower the $SPD$ (better proportional fairness for the suppliers) will be. This indicates the advantage of optimizing for the popularity calibration in the recommendations since it will also benefit the suppliers. 

Another important finding is the fact that more accuracy does not necessarily leads to a better calibration and vice versa as can be seen from Figures ~\ref{fig:upd_prec_movie} and ~\ref{fig:upd_prec_lastfm}. For instance, on Last.fm, \algname{Most-popular} and \algname{Biased-MF} have roughly equal precision but the $UPD$ for \algname{Biased-MF} is significantly lower (better) than \algname{Most-popular}. The reason is, the majority of the items are usually not very popular and, therefore, exclusively recommending popular items would not match the original distribution of the ratings (Figure ~\ref{longtail}) which leads to high $UPD$. In addition, if an algorithm randomly recommends items, the likelihood of having items from $M$ and $T$ increases compared to many personalized recommendations where they suffer from popularity bias. However, improvement in $UPD$ by randomly recommending items would happen under the cost of having an extremely low precision. Therefore, in practice, both $UPD$ and $Precision$ should be taken into account simultaneously in the optimization process.

The relationship between catalog coverage and supplier popularity deviation is also interesting to look at. Will an algorithm that covers many items necessarily have lower (better) $SPD$? The answer is No as can be seen from Figures ~\ref{fig:spd_cov_movie} and ~\ref{fig:spd_cov_lastfm}. The reason is, item coverage (aka aggregate diversity) is the number of unique items that an algorithm has recommended even if an item is only recommended only once. However, what matters in $SPD$ is giving appropriate exposure to the items from different suppliers proportional to their popularity.

The impact of $UPD$ and $SPD$ on the success of a real-world recommender system can be evaluated using online A/B testing and see how metrics such as user engagement, retention, and also the satisfaction of suppliers (e.g. artists in the music recommendation platforms) would be influenced.

% \begin{figure}
% \centering
% \SetFigLayout{2}{1}
%  \subfigure[MovieLens]{\includegraphics[width=2.3in]{Figs/movielens/algs_upd_spd.jpeg}\label{fig:upd_spd_movie}}
%  \subfigure[Last.fm]{\includegraphics[width=2.3in]{Figs/lastfm/algs_upd_spd.jpeg}\label{fig:upd_spd_lastfm}}
%   \hfill
% \caption{Users Popularity propensity Deviation (UPD) vs Supplier Popularity Deviation (SPD)} \label{fig:upd_spd}
% \end{figure} 

% \begin{figure}
% \centering
% \SetFigLayout{2}{1}
%  \subfigure[MovieLens]{\includegraphics[width=2.3in]{Figs/movielens/algs_upd_prec.jpeg}\label{fig:upd_prec_movie}}
%   \hfill
%  \subfigure[Last.fm]{\includegraphics[width=2.3in]{Figs/lastfm/algs_upd_prec.jpeg}\label{fig:upd_prec_lastfm}}
%   \hfill
% \caption{Users Popularity propensity Deviation (UPD) vs Precision} \label{fig:upd_prec}
% \end{figure} 
 
\section{Conclusion}

Recommender systems are multi-stakeholder environments; in addition to the users, some other stakeholders such as the supplier of the items also benefit from the recommendation of their items and gaining a larger audience. The algorithmic popularity bias can negatively impact both users and suppliers on a recommender system platform. In this paper, we demonstrated the severity of the popularity bias impact on different sides of a recommender system using several recommendation algorithms on two datasets. We also proposed metrics to quantify the exposure bias from the perspective of both the users and suppliers. Our experiments showed that when the recommendations are calibrated for the users in terms of popularity (lower $UPD$), it will also benefit the suppliers of the recommendations by giving them proportional exposure (lower $SPD$). We believe, it is extremely crucial for the recommender systems researchers to see the implications of real-world recommenders where a single-stakeholder focus might not address all the complexities.

% \begin{figure}
% \centering
% \SetFigLayout{2}{1}
%  \subfigure[MovieLens]{\includegraphics[width=3in]{Figs/movielens/item_groups.jpeg}}
%   \hfill
%  \subfigure[Last.fm]{\includegraphics[width=3in]{Figs/lastfm/item_groups.jpeg}}
%   \hfill
% \caption{Popularity bias amplification (Item Groups)} \label{ms_impact-Items}
% \end{figure} 

% \input{table2}
\bibliographystyle{ACM-Reference-Format}
\bibliography{acmart}
\end{document}